\documentstyle[12pt]{article}

\begin{document}

 \begin{center}

{\Large \bf ${\cal PT}-$symmetry, ghosts, supersymmetry and Klein-Gordon equation}

\vspace{.5cm}

{ Miloslav Znojil }

  {Nuclear Physics Institute, 250 68 \v{R}e\v{z}, Czech Republic}
%

PACS {03.65.Ge}     

\end{center}

\subsection*{Abstract}

Parallels between the concepts of symmetry, supersymmetry and
(recently introduced) ${\cal PT }-$symmetry are reviewed and
discussed, with particular emphasis on the new insight in quantum
mechanics which is rendered possible by their combined use.

\bigskip

KEYWORDS: {supersymmetric quantum mechanics, parity times time
reversal symmetry, non-Hermitian Hamiltonians with real spectra,
pseudo-metrics in Hilbert space, factorization method,
Klein-Gordon equation} 


\bigskip

\section{Introduction}

When we return to the early stages of development of quantum
mechanics we reveal that the fascination of its authors by their
new discoveries must have been enormous. Suddenly, they were able
to resolve many old and tough puzzles like the ``incomprehensible"
and mysterious stability of atoms with respect to an expected
steady radiation of their moving electrons. Under the name of
${\cal PT}-$symmetry \cite{BB}, perhaps, we might easily be
witnessing a certain continuation of these discoveries in the
nearest future.

As we all know, the ``trick" of the founders of quantum mechanics
was virtually elementary and consisted in a suitable replacement
of any classical observable quantity by an appropriate essentially
self-adjoint operator in Hilbert space ${\cal H}$ (i.e.,
Hamiltonian $H=H^\dagger$ for the energy, etc). Unfortunately,
during the subsequent applications of quantum mechanics, a few
paradoxes emerged in the early forties, especially in connection
with relativistic problems (cf., e.g., the ``Dirac's see"
\cite{see}), representing an important practical limitation of the
whole trick, and hinting that the ``realistic" Hamiltonians might
be non-Hermitian.

This discovery forced the phenomenologically oriented physics
community to forget about the optimism and ambitions of the early
thirties. Theoreticians stampeded towards ``certainties" offered
by the half-explored territory of relativistic quantum field
theory. Such a controversial situation seems to have survived more
than half a century and, admittedly, it forms also an important
part of what we are going to discuss.

In a way inspired by the pioneering letter \cite{BB} we believe
that the generic quantization recipe may be perceived as only too
much innovation-resistent. We are going to review a few aspects of
the Bender's and Boettcher's generalization of the quantization
recipe and of some of its particular applications in the context
of (super)symmetries.

Perhaps, our supplementary motivation lies in the observation that
in its time, the very idea of the quantization was truly
revolutionary. Still, it arose a wave of protests among
conservative physicists of all professional qualities (let us just
mention their most famous EPR branch \cite{EPR}). Hopefully, the
similar protests against ${\cal PT}-$symmetry, however existing
\cite{page}, will prove much less resistent, the more so after the
emergence of its defenses and reviews like the one which follows.

\section{Why ${\cal PT}-$symmetry?}

Only towards the end of the second millennium, Carl Bender dared
to return from field theory to quantum mechanics and formulated
his (nowadays, almost famous) ``but wait a minute!" project
\cite{CBpriv}  where, basically, he advocated the necessity of a
tentative weakening of the (mathematically ``unnecessarily
strong") Hermiticity requirement for the observables. In the
influential letter \cite{BB}, this type of ``heresy" was
formulated as inspired by discussions with their predecessors and
colleagues. In a broader historical perspective, its implicit
origins may be traced back to several independent formal studies
and/or isolated comments reflecting needs of several branches of
phenomenological physics and occurring (or rather ``lost", here
and there) in the more mathematically oriented literature
concerning, mostly, certain peculiar non-Hermitian anharmonic
oscillators with real spectra \cite{Simon}.

It is worth noting at this point that the same idea appeared,
independently, a couple of years sooner, in the fully separate
contexts of nuclear physics (under the name of quasi-Hermiticity
\cite{Geyer}) and in a few other contexts mentioned in \cite{BB}.
Concerning the present state of art, interested reader may
generate easily a comparatively complete set of relevant citations
when looking in the proceedings of the (up to now, two) dedicated
international conferences \cite{CzJ}.

There is probably not yet time for an adequate and critical
evaluation of the resulting  new (though not yet ripe enough)
formulation of the innovated, so called ${\cal PT}-$symmetric
\cite{BB,BBjmp} (or, if you wish, quasi-Hermitian \cite{Geyer}) or
${\cal QPT}-$symmetric \cite{psu} or pseudo-Hermitian \cite{AMA}
or ${\cal CPT}-$symmetric \cite{BBJ,CPT}) quantum mechanics. I may
only note that the use of all these nicknames for the same theory
is, by my own opinion, redundant. That's why I am sticking to the
Bender's terminology (${\cal PT}-$symmetric quantum mechanics,
PTQM), understanding that ${\cal P}$ and ${\cal T}$ need not
necessarily mean just the parity and time reversal, respectively.

In particular, ${\cal T}$ is to be read as an abbreviation for the
more rigorous mathematical requirement of ``being essentially
self-adjoint in ${\cal H}$" \cite{AMC}, so that the time-reversal
symmetry (i.e., the commutativity of the Hamiltonian with the
antilinear operator ${\cal T}$) might find one of its most
frequent applications in the very formal definition of the
standard Hermitian-conjugation mapping, ${\cal T} H {\cal T}
\equiv H^\dagger$.

\subsection{A brief recollection of a few much older relevant works}

In the previous paragraph, we may pre-multiply operator ${\cal T}$
by an indefinite involution ${\cal P}$ and arrive at the explicit
definition of the ${\cal PT}-$symmetry. All the generalized
non-involutory and non-diagonal though, necessarily, non-singular
and essentially self-adjoint \cite{Geyer} forms of our ${\cal P}$
are also admissible and one thus returns just to the old Dirac's
works \cite{Dirac} with his pseudo-metric $\eta$ simply replaced
by~${\cal P}$.

In this spirit, the above-mentioned retreat to field theory was a
pure misunderstandings. Everybody knows that this theory (with all
its infinitely many degrees of freedom and infinitely large
renormalizations etc) represents a not too user-friendly key to
our understanding of microscopic systems. Curiously enough, even
the Dirac-see problem itself, expelled carefully from before the
door of quantum mechanics (and, formally, reducible to an
indefiniteness of the ``physical" Dirac's pseudo-metric $\eta$ in
the full Hilbert space) returned through the window of field
theory. It keeps living there, e.g., as a lasting difficulty with
ghosts (cf. the Pauli's and Gupta's and Bleuler's studies
\cite{Mielnik} and the Lie-Wick's and Nakanishi's discussions
\cite{Nakanishi} or their recent revitalization \cite{Kleefeld}).

\section{How can we survive with indefinite metric?}

The rule $H^\dagger \eta = \eta H$ of pseudo-Hermiticity with a
``Hilbert-space-metric-operator" $\eta = \eta^\dagger \neq
\eta_{traditional} = I$ becomes needed in field theory, it is
possible to say that ``the appearance of negative probability is
the greatest problem in the indefinite metric theory" while ``the
great physicists proposed wrong resolution of it" \cite{Nakapriv}.
A key to the solution of this problem has been offered, very
recently, by Ali Mostafazadeh \cite{AMA} who paid very detailed
attention to one of the most common and physical ${\cal
PT}-$symmetric models, viz., to Klein-Gordon equation in its
Feshbach-Villars form \cite{FV}. He imagined that {\em every}
given and fixed ${\cal PT}-$symmetric Hamiltonian operator $H$
may, in general, satisfy the necessary pseudo-Hermiticity
intertwining rule with {\em many different} pseudo-metrics
$\eta_{m}$. This simple idea (revealed, independently and
practically in parallel, by several groups of other authors
\cite{Geyer,BBJ,ZQ}) initiated a new wave of development of the
theory because some of the new metrics may be positive (i.e.,
$\eta=\eta_+>0$ in the notation of ref. \cite{AMA}).

The first step of all the applications of the new theory should
lie, therefore, in a carefully explained transition from the
``simple" or ``initial" indefinite metric (let us call it ${\cal
P}$) to the ``correct" or rather ``physical" alternative metric
$\eta_+>0$.

A compact denotation $\eta_+ \equiv {\cal CP}$ used in ref.
\cite{BBJ} looks to the present author as one of the best
conventions on the market, with one of his reasons being that the
``charge" operator $C$ coincides with his own (and still older)
quasi-parity $Q$ (cf. \cite{ptho} dating back to 1999). An even
older paper by Scholtz, Geyer and Hahne \cite{Geyer} should be
still more decisively recalled as another recommended reading. In
this reference, the positive $\eta_{+} \neq I$ were also already
known and studied and the related $H$ (be it Hermitian or not in
the usual sense) has been called quasi-Hermitian (this means
Hermitian in the nontrivial, non-isotropic metric ${\cal CP}$)
there.

In such a context, the ghosts of the field theories \cite{ghosts}
emerge as defined as states with the vanishing pseudo-norm defined
with respect to the indefinite metric (i.e., pseudo-metric) ${\cal
P}$. One may then understand the S-matrix as an operator which is
unitary in the sense of the indefinite metric \cite{unit}. It is
not genuinely unitary because the norm positivity is not
guaranteed with respect to ${\cal P}$ \cite{Nakapriv}. Still,
after the change of the metric (i.e., after the re-construction of
the Hamiltonian-dependent quasi-Hermiticity by a re-definition of
the inner product with respect to the positively definite metric
${\cal CP}$) the physical appearance of the negative-norm states
becomes forbidden.

Now, our key message may be formulated as a statement that the
physical interpretation of the ghosts need not necessarily proceed
solely in the traditional Gupta-Bleuler-like ghost-elimination
spirit, i.e., in a way based on an appropriate specification of
the ``physical subspace". The original Hilbert space ${\cal H}$
need not be necessarily declared overcomplete on the physical
grounds, and some subsidiary conditions need not necessarily be
imposed.

Indeed, in the new  metric ${\cal CP}$ (which is to be declared
``correct and physical" and which is, in principle, ambiguous and
dynamically dependent on $H$), one can alternatively get rid of
the interpretation difficulties via the use of the new norm.

\section{Why ${\cal PT}-$symmetric supersymmetry?}

\subsection{Why supersymmetry at all?}

One of the main motivations of the present text originates from
the well known key to the exactly solvable 1D models found in
their supersymmetric (SUSY) re-interpretation \cite{CKS}. It is
amusing to recollect that such an application of SUSY emerged,
historically speaking, as a quite unexpected byproduct of the
originally more ambitious supersymmetric quantum mechanics
(SUSYQM) of Ed Witten.

We consider the latter formalism worth extending to the ${\cal
PT}-$symmetric world. As long as the ${\cal PT}-$symmetry property
itself is not too dissimilar from its ``Hermiticity" predecessor
$H = H^\dagger$, one may expect that an active use of ${\cal
PT}-$symmetric models could be capable of altering the present
status and the role of all the standard symmetries in general and
of the supersymmetry in particular.


To set the scene, let us return to the sample linear parity
operator ${\cal P}$ and to an eigenvalue problem $H\,|\Psi\rangle
= E\,|\Psi\rangle$ with symmetry $H\,{\cal P} = {\cal P}\,H$. In
the light of Schur's lemma this implies that every linear
combination of $|\Psi\rangle$ and ${\cal P}\,|\Psi\rangle$ will
also satisfy the same eigenvalue problem so that in the most
common one-dimensional and non-degenerate (= Sturm-Liouville)
setting we may immediately classify all the solutions
$|\Psi\rangle$ according to their parity.

Once we move to the more sophisticated symmetries, the same
procedure makes our spectra multiply indexed. One of the most
successful applications of such a strategy may be undoubtedly
found in the physics of elementary particles where the symmetries
of the interactions proved to be a powerful source of the
classification of the possible solutions (= particle multiplets).

Paradoxically enough, SUSY as a mathematically most natural
transition to the symmetries between the bosons and fermions
(called, in mathematics, graded algebras) failed in practice. No
SUSY-partner element of any supermultiplet has been found up to
now. At the same time, the use of the first nontrivial graded Lie
algebra $sl(1|1)$ proved extremely fruitful within SUSYQM. In
Hermitian case, its three 'graded' generators (viz., Hamiltonian
$H$ and the two ``supercharges $Q$ and $P$ satisfying there the
'fermionic' nilpotence rule $PP=QQ=0$ plus a compatibility
commutation relations $ H P - P H = H Q - Q H = 0$) are described
in detail, say, in review paper \cite{CKS}.

\subsection{Why supersymmetry without Hermiticity?}

An explicit sample of the modified SUSYQM in its various ${\cal
PT}-$symmetric versions may be found among papers \cite{CZ}.
Probably the first application of such a non-Hermitian
quantum-mechanical SUSY formalism to the fully and exactly
solvable (spiked) harmonic oscillator  may be found in ref.
\cite{HOZn}. At this point it would be useful to carbon-copy some
formulae for illustration but we must skip such a plan due to the
absolute shortage of space for such a purpose.

In the middle of our very concise discussion of the subject,
surviving without the formulae, it is still necessary to emphasize
that ${\cal PT}-$symmetry is anti-linear (i.e., non-linear) so
that all the apparent Schur-type parallels with the ordinary
linear symmetries (and SUSY) will immediately break down
\cite{CBpriv}. Similar formal challenges accelerated, after all,
the very recent development of the field. I.a., it has been found
that a much easier access to many relevant structures immanent in
the majority of the ``weakly Hermitian" ${\cal PT}-$symmetric
models (and, in particular, to their specific spectral
representations \cite{ZQ,AM} and/or to the existence of the very
specific, so called exceptional points in their spectra
\cite{Heiss}) may be mediated by many much more elementary models
\cite{ostatni}.

This further supported our interest in SUSY constructions in a way
which applies, in particular, to the really exceptional exactly
solvable harmonic-oscillator limit of virtually all of the
above-mentioned peculiar anharmonic oscillators \cite{ptho}. In
the same direction, the first successful steps have been made also
towards the more-particle ${\cal PT}-$symmetric exactly-solvable
models of the Calogero \cite{Tater} or Winternitz \cite{Jakub}
types. In this context, many questions still remain open
\cite{JakuBiDipl}.

\section{${\cal PT}$ form of SUSY in application to Klein-Gordon
equation}

In an overall non-Hermitian setting, the SUSY Hamiltonian $H$
becomes a direct sum of its diagonal 'left' and 'right'
Hamiltonian-type sub-operators $H_{(L,R)}$. In the same two-by-two
partitioned notation (cf. \cite{CKS} for all details) both $Q$ and
$P$ are, respectively, lower and upper triangular two-by-two
matrices. They contain just one off-diagonal element (say,
operators $a$ and $c$, respectively). Of course, the standard
Hilbert-space representations of the latter $a$ (annihilation
operator) and $c$ (creation operator) are usually  non-diagonal
and may be often written in the one-diagonal upper- and
lower-triangular infinite-dimensional matrix form, respectively.

In our recent paper \cite{Vallad} we asked what happens if one
relaxes the standard Hermiticity requirements. What we did in ref.
\cite{Vallad} was, in essence, just a transfer of the underlying
SUSY-type factorization of the Hamiltonian to the domain of the
Klein-Gordon-type equations. Interested reader is recommended to
search for the explicit formulae in {\it loc. cit.}.

Basically, we proceeded in full analogy with the non-relativistic
case. Keeping ${\cal P}$ equal to the (most elementary operator
of) parity we recollected the standard procedure (seen as a source
of interest in the imaginary cubic anharmonicities $-ix^3$ in the
late seventies) and assumed that any suitable preselected
spatially symmetric and real (read: ${\cal T}-$symmetric)
potential is made non-Hermitian and ${\cal PT}-$symmetric by
adding another, purely imaginary and spatially antisymmetric
component to it. A deeper understanding of the similar models can
already rely on the intensive technical developments in the field.
Thus, explicit constructions may be facilitated by a recourse to
the delta-expansion techniques of field theory \cite{BM} or to the
WKB \cite{WKB} and strong-coupling \cite{FRGZ} perturbation
expansions or to the quasi-exact solution techniques \cite{BB2}.
Last but not least, the use of the language of Bethe ansatz might
prove also an enormously efficient tool \cite{DDT}.

\section{Conclusions}

\subsection{Practical use of ${\cal PT}-$symmetry}

We emphasized that once we accept the necessity of a Lorentz
covariance of some realistic dynamical equations in quantum
setting, we arrive at one of the oldest, most physical and most
popular PTQM example representing the Klein-Gordon version of the
relativistic quantum mechanics in the form authored by Feshbach
and Villars in the middle of fifties \cite{FV},  clarified to be
consistent with the standard postulates of quantum mechanics by
Ali Mostafazadeh \cite{AMG} and encountering an unexpected
resurrection in cosmology at present \cite{AMW}.

In this model, a successful combination of the Lie (=linear)
symmetry with the non-Hermiticity of the generator $H_{FV}$ of the
time evolution exemplifies a specific non-diagonal and two-by-two
partitioned form of the indefinite metric ${\cal P}$ or $\eta$.


Classical representation theory did find a place for both the
linear and antilinear operators (cf. reviews of this topic
\cite{SW}). The related analysis of the Lie-algebraic background
of the exact solvability of the one-dimensional Schr\"{o}dinger
equations has been also recently extended to the case of the
${\cal PT}-$symmetric models by Bagchi and Quesne who revealed
that a weakening of the Hermiticity requirement implies that the
solvable models will form a broader class exhibiting, i.a., an
enrichment of their symmetry algebras by complexifying the
standard $so(2,1)$ to $sl(2, C)$ \cite{Quesne}.

In the Lie-algebraic setting, a particularly useful role seems to
be played by the particular Calogero-type models which mimic a
realistic multiparticle dynamics in one dimension. One might note
that a suitable ${\cal PT}-$symmetric complexification of these
models has been shown also to open a path towards a nonstandard
limiting transition to the entirely new solvable models
\cite{Taterlett}.

\subsection{${\cal PT}-$symmetry in
combination with SUSY}

It is rather amusing to note \cite{Canpriv} that the 1998 letter
\cite{CJT} (which might be thought of as one of the very first
texts on ${\cal PT}-$symmetry in SUSY systems) paid its attention
to only too many features of the problem at once so that, e.g.,
the exactly solvable model it describes in its last chapter seems
to be almost forgotten at present.

One of explanations is that both its second and third authors have
already left physics completely. In contrast, its first author
remains extremely active in the field and should be acknowledged
for having introduced the author of this lecture in the field in
1999.  Unfortunately, once this summary of my Prague's lecture has
a very restricted number of pages, the lack of space forced me to
skip virtually all the technical details of my own papers, and the
more so in the case of many relevant and extremely interesting
results produced, e.g., by F. Cannata and his co-authors
\cite{Cannata}, by C. Quesne and her co-authors \cite{Quesneb}
etc.

\subsection{Outlook}

Let us summarize that when working simply with the two alternative
metric operators, one of them may remain indeterminate. For the
purely practical purposes it is only necessary that its structure
is ``sufficiently simple" (typical examples: the Bender's parity
$P$ or the Feshbach-Villars' $\sigma_3$). The second metric
should then be constructed as positively definite, allowing us to
define the norms of states. One can hardly expect that the latter
operator would be not too complicated.

The progress in the whole theory of this type is marked by
explicit examples of the desirable physical metrics. Besides the
early product $QP$ (assigned to the exactly solvable ${\cal
PT}-$symmetric spiked harmonic oscillator in 1999 \cite{ptho}),
one should not forget the Ali Mostafazadeh's alternative to the
Feshbach-Villars'  metric for the Klein-Gordon field \cite{AMG}
and, last but not least, several fresh, beautiful and explicit
constructions of the products $CP$ obtained by different
sophisticated methods for several different field models by Carl
Bender and his co-authors \cite{Wang}.

On this background, the standard studies of SUSY models are also
re-acquiring a new motivation. The present review mentioned the
few steps in this direction, exhibiting already certain clear
parallels between pseudo- and Hermitian SUSY. By the present
author's opinion, these preliminary sample results just mark a
very start of a more intensive development of this subject in some
very near future.

\bigskip
 \noindent
{\small {\bf Acknowledgements:} (Partially emailed) recent
discussions with F. Kleefeld and N. Nakanishi are particularly
appreciated. Work was partially supported by the grant number A
1048302 of GA AS CR.}

\end {document}